\documentclass[a4paper,fleqn]{cas-dc}

\usepackage[numbers,sort&compress]{natbib}

\usepackage[english]{babel}

\usepackage{graphics}
\usepackage{xcolor}
\usepackage{amssymb}
\usepackage{amsmath}
\usepackage{overpic}
\usepackage{epstopdf}
\usepackage{threeparttable}
\usepackage{tabularx}
\usepackage{adjustbox}
\usepackage{booktabs}
\usepackage{lipsum}
\usepackage{braket}

\usepackage{bm}
\usepackage{graphicx}
\usepackage{subfigure}
\usepackage{multirow}
\usepackage{array}
\usepackage{newtxtext}
\usepackage{siunitx}

\usepackage{hyperref}
\hypersetup{plainpages=false,colorlinks=true,linkcolor=blue, citecolor=blue, urlcolor=blue}

\def\tsc#1{\csdef{#1}{\textsc{\lowercase{#1}}\xspace}}
\tsc{WGM}
\tsc{QE}

\begin{document}
\let\WriteBookmarks\relax
\def\floatpagepagefraction{1}
\def\textpagefraction{.001}




\title[mode = title]{Chemical-substitution-driven
giant anomalous Hall and Nernst effects in magnetic cubic Heusler compounds}


%


\author[1]{Guangzong Xing}
                        [
                        orcid=0000-0002-8299-8585]
\cormark[1]
\ead{XING.Guangzong@nims.go.jp}


\address[1]{Research Center for Magnetic and Spintronic Materials, National Institute for Materials Science, Tsukuba, Ibaraki, 305-0047, Japan}
\address[2]{Digital Transformation Initiative Center for Magnetic Materials, National Institute for Materials Science, Tsukuba, Ibaraki, 305-0047, Japan}

\author[1,2]{Keisuke Masuda}[orcid=0000-0002-6884-6390]
\cormark[1]



\ead{MASUDA.Keisuke@nims.go.jp}


\author[1,2]{Terumasa Tadano}
[orcid = 0000-0002-8132-2161 ]


\author[1,2]{Yoshio Miura}
[
   orcid = 0000-0002-5605-5452]


\begin{abstract}
 Chemical substitution efficiently optimizes the physical properties of Heusler compounds, especially their anomalous transport properties, including anomalous Hall conductivity (AHC) and anomalous Nernst conductivity (ANC). This study systematically investigates the effect of chemical substitution on AHC and ANC in 1493 magnetic cubic Heusler compounds using high-throughput first-principles calculations. 
  Notable trends emerge in Co- and Rh-based compounds, where chemical substitution effectively enhances the AHC and ANC. Intriguingly, certain chemically substituted candidates exhibit outstanding enhancement in AHCs and ANCs, such as (Co$_{0.8}$Ni$_{0.2}$)$_2$FeSn with considerable AHC and ANC values of $-2567.78$~S\,cm$^{-1}$ and $8.27$~A\,m$^{-1}$K$^{-1}$, respectively, and (Rh$_{0.8}$Ru$_{0.2}$)$_2$MnIn with an AHC of $1950.49$~S\,cm$^{-1}$. In particular, an extraordinary ANC of $8.57$~A\,m$^{-1}$K$^{-1}$ is identified exclusively in Rh$_2$Co$_{0.7}$Fe$_{0.3}$In, nearly double the maximum value of $4.36$~A\,m$^{-1}$K$^{-1}$ observed in the stoichiometric Rh$_2$CoIn. A comprehensive band structure analysis underscores that the notable enhancement in ANC 
  arises from the 
  creation and modification of the energy-dependent nodal lines through chemical substitution. 
  This mechanism generates a robust Berry curvature, resulting in significant ANCs. These findings emphasize the pivotal role of chemical substitution in engineering high-performance materials, thereby
  expanding the horizons of transport property optimization within Heusler compounds.    
\end{abstract}




\maketitle

\section{Introduction}

Ever since the discovery of the Seebeck effect in 1826~\cite{Seebeck}, considerable efforts have been directed toward improving the performance of conventional thermoelectric devices. The Seebeck effect, known as the 
longitudinal thermoelectric effect, involves the generation of an electric field parallel to the applied temperature gradient. In general, complex device structures are typically required to enhance thermoelectric performance. In recent years, transverse thermoelectric devices driven by the anomalous Nernst effect (ANE) have gained increased interest due to their inherent simplicity~\cite{uchida2022anisotropy,uchida2021transverse}.

The ANE serves as the thermoelectric counterpart to the anomalous Hall effect (AHE), wherein a transverse electric field is generated when a perpendicular temperature gradient is applied in the absence of the external magnetic filed~\cite{Nernst}. Similarly, the AHE describes the generation of a transverse electric field in response to the electric field along the perpendicular direction~\cite{nagaosa2010anomalous}. The linear response coefficients of these effects, known as the anomalous Hall conductivity (AHC) and anomalous Nernst conductivity (ANC), originate from two primary contributions: extrinsic and intrinsic. Extrinsic contributions are related to the electron scattering by impurities, whereas the intrinsic component of AHC and ANC purely originates from the electronic structure, driven by the summation of the Berry curvature (BC) within the first Brillouin zone (FBZ)~\cite{xiao2005berry,yao2004first}. These anomalous transport behaviors are commonly observed in magnetic materials, where a robust BC is induced through breaking the time-reversal symmetry, driven by the Weyl points or nodal lines (NL) with nontrivial topological characteristics~\cite{sakai2020iron,pan2022giant,liu2019magnetic,chen2021anomalous,kim2018large,suzuki2016large,he2022topological,fu2020topological}.

Among various categories of magnetic materials, Heusler compounds of $L2_1$ type are of great interest. They exhibit promising properties for various applications, such as high spin polarization~\cite{kurniawan2021first}, high Curie temperature~\cite{HU2023119255}, novel shape-memory~\cite{PhysRevLett.104.176401,TAKHSHAGHAHFAROKHI2023118603} effect and notable magnetocaloric effect~\cite{taubel2020tailoring}. 
Remarkably, the intriguing anomalous transport properties of these materials have been extensively investigated by both experimental analysis and first-principles calculations based on density functional theory (DFT)~\cite{manna2018colossal,sumida2020spin,yang2023intrinsic}. 
Theoretical work by Noky~$et$~$al.$~\cite{noky2019large,noky2018characterization} focused on Co- and Fe-based Heusler compounds and revealed the vital role of  
mirror symmetry-protected NLs in generating substantial AHC and ANC. 
It is important to note that the mirror symmetry of $L2_1$-type Heusler compounds causes the sum of BCs induced by gapless NLs to cancel out, resulting in zero AHC and ANC.
However, the NL becomes gapped when mirror symmetry is broken by a small perturbation such as the spin-orbit coupling (SOC). Consequently, strong BC is induced in the gapped NLs, leading to a large AHC and ANC. For instance, in the ferromagnetic Co$_2$MnGa, several research groups have reported a large ANC of $\sim$$4$~A\,m$^{-1}$K$^{-1}$ 
at $300$~K, accompanied by a considerable AHC with the magnitude of $10^3$~S\,cm$^{-1}$~\cite{belopolski2019discovery,suzuki2016large,guin2019anomalous}. The presence of NL and Weyl points near the Fermi energy in Co$_2$MnGa was confirmed by the angle-resolved photoemission spectroscopy and the first-principles calculations~\cite{suzuki2016large,belopolski2019discovery,sumida2020spin,minami2020enhancement}. 

Theoretical investigation of anomalous transport properties has focused mainly on stoichiometric materials. 
However, it is important to recognize that chemical substitution offers a promising way to tailor these properties, and the effectiveness of enhancing AHC and ANC through chemical substitution has been observed in various experimental studies.
Shen~$et$~$al.$~\cite{shen2020local} reported an enhanced AHC in (Co,Ni)$_3$Sn$_2$S$_2$ alloy. Increased ANC was observed in Co$_2$Mn(Al,Si) and Co$_2$(Ti,V)Sn alloys~\cite{sakuraba2020giant,hu2018anomalous}. Previous theoretical studies have mainly focused on stoichiometric compounds and estimated the effect of chemical substitution using the band-filling approach~\cite{tanzim2023giant}, which fails to encompass electronic structure changes due to chemical substitution.  
To directly understand the chemical substitution effect on anomalous transport properties of a wide range of compounds, employing first-principle high-throughput calculations becomes imperative. Noky~$et$~$al.$~\cite{noky2020giant} performed a comprehensive study on the anomalous transport properties in magnetic cubic Heusler compounds and identified potential candidates with significant AHC and ANC. However, their study did not directly consider the influence of chemical substitution on AHC and ANC. 
To date, there is a notable absence of comprehensive studies that directly address the effect of chemical substitution on the anomalous transport properties in Heusler compounds.

In response, a comprehensive high-throughput approach is employed in our study to directly explore the effect of chemical substitution on AHC and ANC in Heusler compounds. 
The AHC and ANC are computed using the tight-binding Hamiltonian by automatically constructing the 
maximally localized Wannier functions. Our study encompasses a vast data set of $1493$ compounds derived from $104$ stoichiometric mother compounds with potential AHC or ANC, as reported by Noky~$et$~$al.$\cite{noky2020giant}. 
Chemical substitution is treated using virtual crystal approximation (VCA), a conventional method commonly employed for various properties in magnetic materials~\cite{bellaiche2000virtual,yanagi2021first,zhang2011anisotropic,huang2015anomalous}. Within the VCA framework, a virtual element, denoted as $X_{1-x}X_x^\prime$, is employed, where $X$ and $X^\prime$ represent the original element and its substitution, respectively, with $x$ indicating the concentration of the substitution. This approach ensures the preservation of the symmetry between the investigated candidates and their corresponding stoichiometric counterparts. Furthermore, the justification for the use of VCA to maintain symmetry is supported by experimental observations of the $L2_1$ order in off-stoichiometric Co$_2$MnAl, confirmed by x-ray fluorescence analysis in certain cases of chemical substitution with Si~\cite{sakuraba2020giant}. Our study identified a substantial number of Heusler compounds with chemical substitution that exhibited enhanced AHC and ANC. Notably, a series of chemically substituted compounds showcase ANC values at the Fermi energy surpassing peak values in their respective stoichiometric mother compounds. This originates from a significant change in the electronic structure due to chemical substitution. 
Of particular interest is the nearly twofold increase in ANC in Rh$_{2}$Co$_{0.7}$Fe$_{0.3}$In, reaching $8.57$~A\,m$^{-1}$K$^{-1}$ at $300$~K, compared to the peak value in the Rh$_2$CoIn ($4.36$~A\,m$^{-1}$K$^{-1}$). The band structure analysis underscores the crucial role of chemical substitution in creating or modifying the NLs, ultimately leading to significant enhancements in both AHC and ANC.

\section{Computational method}
\label{sec:method}
\subsection{Description of the workflow}
\label{subsec:workflow}

\begin{figure*}
   \centering
   \includegraphics[width=0.95\textwidth]{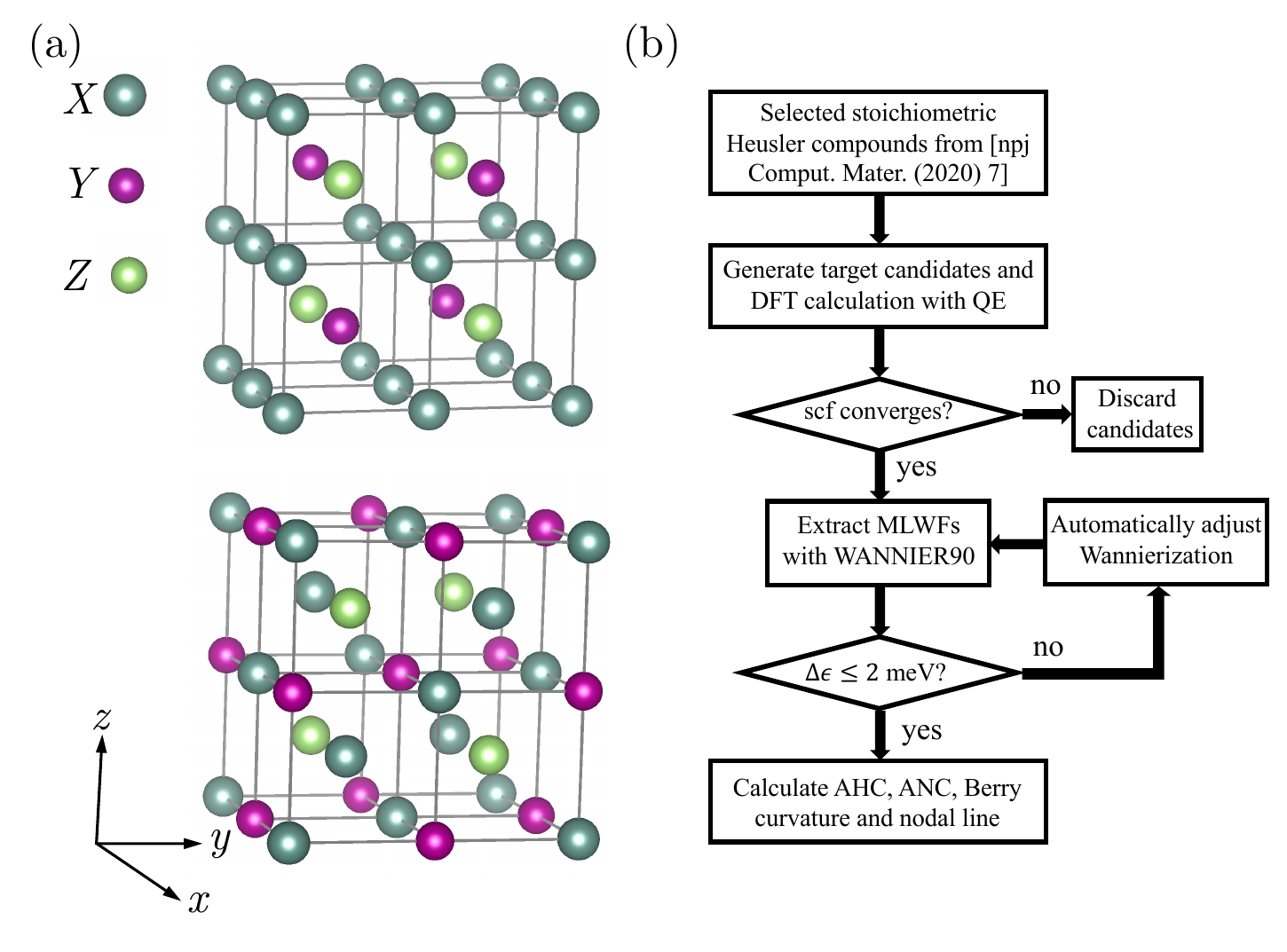}
   \caption{(a) Crystal structure of the $L2_1$-type (upper panel) and inverse (lower panel)
  Heusler compounds, respectively. (b) Workflow for the present high-throughput screening. The initial stoichiometric mother compounds were selected from Ref.~\cite{noky2020giant} based on the maximum values of 
  $\sigma_{xy}$ or $\alpha_{xy}$, which are larger than or equal to 700~S cm$^{-1}$ and 3~A m$^{-1}$K$^{-1}$, respectively.}
   \label{fig:Fig1}
  \end{figure*}

Figure~\ref{fig:Fig1}(b) illustrates the workflow of the high-throughput calculation, where transport properties were automatically computed using the in-house-developed Python scheme linked to \textsc{Quantum Espresso} (QE),  \textsc{Wannier90}, and \textsc{Wanniertools} software. The initial stoichiometric compounds were selected from the database of Noky $et$ $al.$~\cite{noky2020giant} based on the maximum of AHC or ANC ($T=300$~K) in the energy window of $0.25$ eV around the Fermi energy, ensuring a threshold of $700$~S\,cm$^{-1}$ and $3$~A\,m$^{-1}$K$^{-1}$, respectively. In total, 104 compounds were selected, comprising $98$ $L2_1$-type (space group $Fm\bar{3}m$, \# 225) and $6$ inverse (space group $F\bar{4}3m$, \# 216) compounds, [see Fig.~\ref{fig:Fig1}(a)] respectively.  Three types of Heusler compounds with chemical substitution, characterized by the chemical formula $(X_{1-x}X^\prime_x)_2YZ$, $X_2Y_{1-x}{Y^\prime}_{x}Z$ and $X_2YZ_{1-x}{Z^\prime}_x$, were simulated using VCA, where $X^\prime$ ($Y^\prime$, $Z^\prime$) is the neighbor element of $X$ ($Y$, $Z$) and $x= 0$, $0.1$, $0.2$, $0.3$, respectively. In $L2_{1}$-type Heusler structures, the $X_{1-x}X_x^\prime$, $Y_{1-x}Y_x^\prime$, and $Z_{1-x}Z_x^\prime$ atoms occupy the Wyckoff positions $8c$ ($1/4$, $1/4$, $1/4$), $4a$ ($0$, $0$, $0$), and $4b$ ($1/2$, $1/2$, $1/2$), respectively. However, in the inverse Heusler structures, half of the $8c$ $X_{1-x}X_x^\prime$ atoms are replaced by the $4a$ $Y_{1-x}Y_x^\prime$ atom, and the Wyckoff positions are redefined as $X_{1-x}X_x^\prime$ [$4a$ ($0$, $0$, $0$)], $X_{1-x}X_{x}^\prime$ [$4c$ ($1/4$, $1/4$, $1/4$)], $Y_{1-x}Y_{x}^\prime$ [$4d$ ($3/4$, $3/4$, $3/4$)], and $Z_{1-x}Z_{x}^\prime$ [$4b$ ($1/2$, $1/2$, $1/2$)], respectively. A total of $1528$ candidates were generated as input for this comprehensive study. 

Here, all investigated candidates were considered to be in collinear ferromagnetic or ferrimagnetic states, and as such, the initial magnetization was aligned along the [001] Cartesian crystal axis.
The criteria used to assess the energy difference, denoted as
\begin{align}
   \Delta{\epsilon}=\sqrt{\frac{1}{N_{n,\bm{k}}}\sum_{n,\bm{k}}(\epsilon_{n,\bm{k}}^{\mathrm{DFT}}-\epsilon_{n,\bm{k}}^{\mathrm{Wan}})^2} 
\end{align}
where $\epsilon_{n,\bm{k}}^{\mathrm{DFT}}$ and $\epsilon_{n,\bm{k}}^{\mathrm{Wan}}$ denote the eigenenergy obtained by QE and \textsc{Wannier90} codes, respectively. $n$ and $\bm{k}$ represent the band index and Bloch wave vector, respectively. The energy of $\epsilon_{n,\bm{k}}^{\mathrm{DFT}}$ ($\epsilon_{n,\bm{k}}^{\mathrm{Wan}}$) obtaining $\Delta\epsilon$ is constrained to be lower than $\epsilon_{\mathrm{F}}+1$~eV, where $\epsilon_{\mathrm{F}}$ represents the Fermi energy. $N_{n,\bm{k}}$ is the total number of points at each $n$ and $\bm{k}$ within the defined energy range.  The largest $\Delta\epsilon$ of $2.02$~meV was observed in Fe$_2$ScAl$_{0.9}$Si$_{0.1}$. As shown in Fig.~S1 of the Supplemental Material (SM), an almost perfect match of the band structure obtained from DFT calculations and Wannierization suggests that the applied threshold of $\sim$$2$~meV is reasonable to obtain accurate AHCs and ANCs. 
Subsequently, all candidates were adopted to the workflow and the AHC and ANC were calculated if the aforementioned criteria were met. As a result, a success rate of $\sim$98\% (1493 candidates) was 
achieved in this investigation.

\subsection{Intrinsic contribution to the AHC and ANC}
\label{subsec:ahc}
The second-rank antisymmetric AHC tensor can be described as a pseudovector $\bm{\sigma}=(\sigma_{yz},\sigma_{xz},\sigma_{xy})$, whose shape is associated with the direction of the magnetization. All examined candidates, including $L2_1$-type and inverse Heusler compounds, belong to the magnetic Laue group $4/mm'm'$, where prime denotes time-reversal symmetry, when the magnetization is oriented along the [001] direction~\cite{aroyo2011crystallography,aroyo2006bilbao,aroyo2006bilbaoII}. Constrained by the pseudovector transformation under the symmetry,
only $\sigma_{xy}$ remains nontrivial in this context~\cite{seemann2015symmetry,park2022anomalous,tanzim2023giant}.

The intrinsic component of anomalous Hall conductivity $\sigma_{xy}$, which arises solely from the electronic
band structure, can be directly evaluated using the Kubo
formula~\cite{xiao2010berry,nagaosa2010anomalous},
\begin{align}
   \sigma_{xy} & = -\frac{e^2}{\hbar}\int_{\mathrm{BZ}}\frac{d\bm{k}}{(2\pi)^3}\sum_n{}f(\epsilon_{n\bm{k}})\Omega_{n,xy}(\bm{k}).
   \label{eq:AHC}
\end{align}
Here, $\hbar$, $e$, and $\epsilon_{n\bm{k}}$ represent the reduced Planck constant, positive elementary charge, and eigenenergy.
The Fermi distribution function is denoted by $f(\epsilon)=(\mathrm{e}^{(\epsilon-\mu)/k_{\mathrm{B}}T}+1)^{-1}$, where $\mu$ stands for the chemical potential. The Berry curvature $\Omega_{n,xy}(\bm{k})$ for band $n$, which can be expressed as,  
\begin{align}
   \Omega_{n,xy}(\bm{k})=-2\hbar^2\mathrm{Im}\sum_{m(\neq{}n)}\frac{\braket{n\bm{k}|\hat{v}_x|m\bm{k}}\braket{m\bm{k}|\hat{v}_y|n\bm{k}}}{(\epsilon_{n\bm{k}}-\epsilon_{m\bm{k}})^2},
   \label{eq:BC}
\end{align}
with $\hat{v}_x$ ($\hat{v}_y$) being the $k_x$ ($k_y$) component of the velocity operator, and $\ket{n\bm{k}}$ representing the eigenstate.

Similarly, the anomalous Nernst conductivity $\alpha_{xy}$ can be obtained by integrating the BC over the first Brillouin zone with a different 
occupation function. It can be expressed as~\cite{ghimire2019creating,yanagi2021first},
\begin{align}
   \alpha_{xy} & = \frac{ek_{\mathrm{B}}}{\hbar}\int_{\mathrm{BZ}}\frac{d\bm{k}}{(2\pi)^3}\sum_ns(\epsilon_{n\bm{k}})\Omega_{n,xy}(\bm{k}),
   \label{eq:ANC}
   \end{align}
where $s(\epsilon)=-f(\epsilon)\mathrm{ln}f(\epsilon)-[1-f(\epsilon)]\mathrm{ln}[1-f(\epsilon)]$ is the density entropy function, and $k_{\mathrm{B}}$ denotes the Boltzmann constant. As proposed by Xiao $et$ $al.$, $\alpha_{xy}$ can be connected with $\sigma_{xy}$ through integration by parts, which is written as~\cite{xiao2006berry},
\begin{align}
   \alpha_{xy} = \frac{1}{eT}\int{d\epsilon}~(\epsilon-\mu)\frac{\partial{f}}{\partial{\epsilon}}\sigma_{xy}(\epsilon,T=0)
   \label{eq:ANC1}
\end{align}

\subsection{Computational details}
\label{subsec:detail}

We perform first-principles calculations of the electronic band structure using the \textsc{Quantum} ESPRESSO code~\cite{Giannozzi_2009,giannozzi2017advanced}, where the generalized gradient approximation (GGA), proposed by Perdew, Burke, and Ernzerhof (PBE)~\cite{perdew1996generalized}, was adopted. The fully relativistic optimized normal-conserving Vanderbilt (ONCV) peudopotentials~\cite{hamann2013optimized} 
taken from the PseudoDojo~\cite{van2018pseudodojo}  were employed. Spin-orbit coupling was included for all transport quantity and BC calculations. A kinetic cutoff of 100 and 400~Ry was chosen for the wave function and charge density, respectively. For self-consistent calculation, a $k$-point mesh density of $\sim$450~\si{\angstrom}$^3$ was employed, with a total energy convergence criterion of 10$^{-10}$ Ry. 
We adopted the same lattice parameters as those reported in Ref.~\cite{noky2020giant} for both the chemically substituted Heusler alloys and their corresponding stoichiometric counterparts to compare the results.

\begin{figure*}
   \centering
   \includegraphics[width=0.95\textwidth]{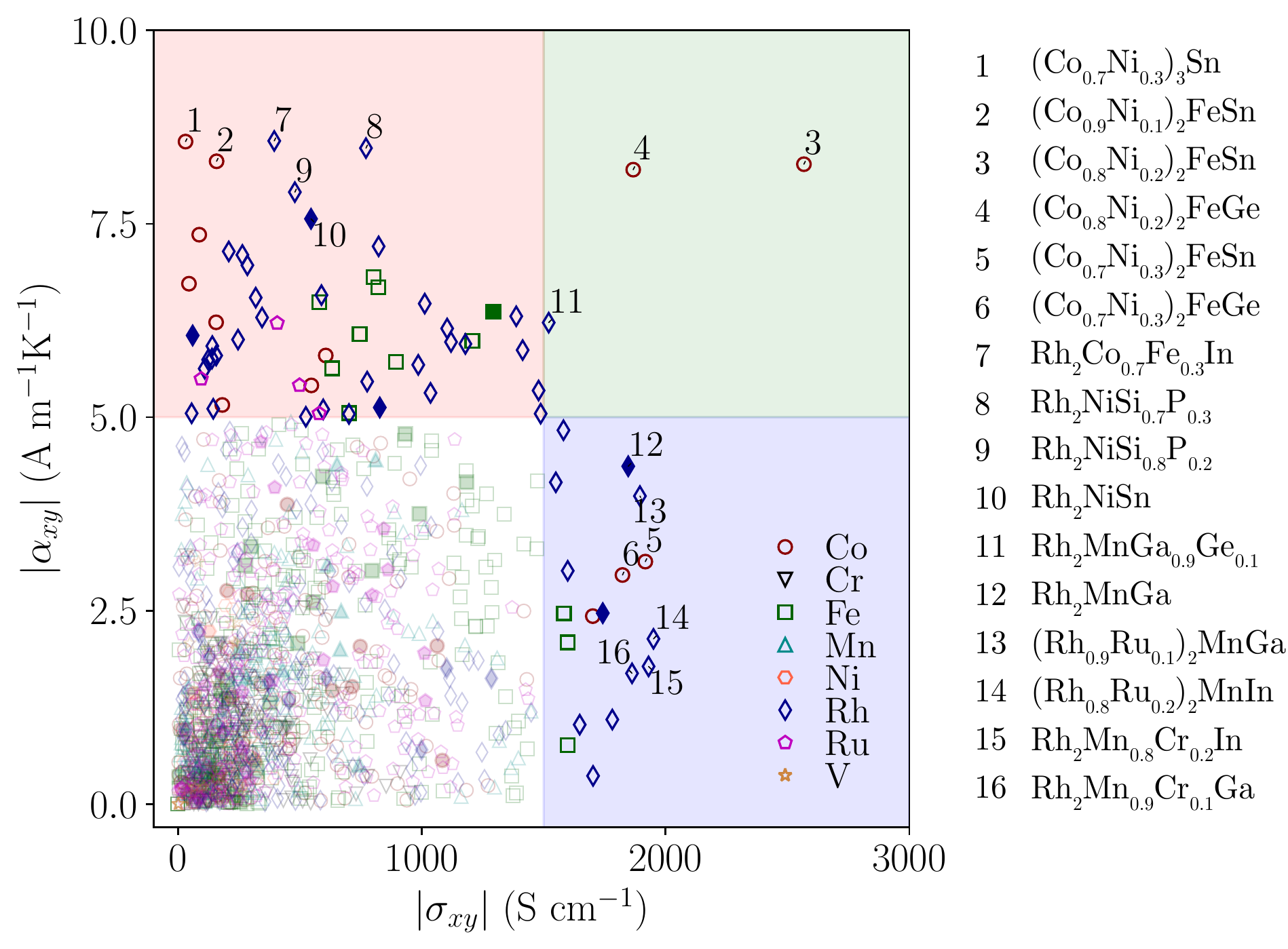}
   \caption{Absolute value of $\sigma_{xy}$ (at $T=0$~K) and $\alpha_{xy}$ (at $T=300$~K) for investigated Heusler compounds (1493 in total).
  Filled and open symbols with the same color represent stoichiometric and chemically substituted compounds containing the same element at the $X$ site.
  The red and blue areas indicate candidates with large $|\sigma_{xy}|$ and $|\alpha_{xy}|$, respectively, while the green area 
  involves the candidates that possess both large $|\sigma_{xy}|$ and $|\alpha_{xy}|$.} 
   \label{fig:Fig2}
  \end{figure*}

The $\sigma_{xy}$ was calculated based on the maximally localized Wannier function (MLWF) using the \textsc{Wannier90} code~\cite{marzari1997maximally,souza2001maximally}. The tight-binding Hamiltonian was obtained
from first-principles calculations with a uniform $10\times10\times10$ $k$-mesh. To construct the MLWF, we employ the selected columns of the density matrix (SCDM) 
method~\cite{vitale2020automated,damle2015compressed,damle2018disentanglement}, which is well-suited for high-throughput calculations. 
SCDM requires only two free parameters, automatically determined in our Python scheme. The outer energy window is automatically adjusted in the disentanglement procedure until $\Delta\epsilon$ reaches the defined criterion. For detailed parameters of Wannierization, please refer to the SM. 
The $\sigma_{xy}$ was computed using a $150\times150\times150$ $k$-mesh in the FBZ.
A denser $k$-mesh of $200\times200\times200$ was used for Co$_2$MnGa and a change of $\sigma_{xy}$ within ~$\sim$1\% was observed, suggesting that the current $k$-mesh was dense
enough. $\alpha_{xy}$ was evaluated using
Equation~(\ref{eq:ANC1}) within the energy range of $[-0.4{:}0.4]$~eV with respect to the Fermi energy at 300~K. The BC and nodal line network of the selected candidates were analyzed using \textsc{wanniertools} software~\cite{WU2017}. 

\section{Results and discussion}

\begin{table*}
   \begin{adjustbox}{max width = 2\columnwidth}
   \begin{threeparttable}
      \centering
      \small
      \caption{Identified top 10 promising $L2_1$-type candidates with large $|\sigma_{xy}|$ (at $T=0$~K) at the Fermi energy.}
         \label{table:sig}
         \renewcommand{\arraystretch}{1.25}
         \begin{tabular}{lccccccc}
         \hline\hline
         Candidates &SG & $m$ & $\sigma_{xy}$ & $\alpha_{xy}$ & $\sigma_{xy}^{\mathrm{max}}$ [$\Delta\epsilon$] &
         $\alpha_{xy}^{\mathrm{max}}$ [$\Delta\epsilon$] \\
                 & &($\mu_{\mathrm{B}}$/f.u.) & (S\,cm$^{-1}$) & (A\,m$^{-1}$K$^{-1}$)  &  (S\,cm$^{-1}$) [(eV)]& (A\,m$^{-1}$K$^{-1}$) [(eV)] \\
         \toprule[0.15mm]
         (Co$_{0.8}$Ni$_{0.2}$)$_2$FeSn & $225$ & $5.33$ & $-2567.78$ & $8.27$ & $-3224.09$ ($0.02$) & $11.99$ ($-0.03$) \\
         (Rh$_{0.8}$Ru$_{0.2}$)$_2$MnIn & $225$ & $3.95$ & $1950.49$ & $-2.14$ &$1950.49$ ($0.00$) & $-6.67$ ($-0.05$)\\
         (Rh$_{0.9}$Ru$_{0.1}$)$_2$MnGa & $225$ & $3.94$ & $1894.95$ & $-3.98$ &$2148.66$ ($0.04$) & $-7.09$ ($-0.07$)\\
         (Co$_{0.8}$Ni$_{0.2}$)$_2$FeGe & $225$ & $5.31$ & $-1867.45$ & $8.20$ &$-3152.20$ ($0.02$)& $10.90$ ($-0.03$) \\
         Rh$_2$MnAl & $225$ & $4.11$, $4.1^{\mathrm{a}}$ & $1742.50$, $1723^{\mathrm{a}}$ & $2.47$, $2.26^{\mathrm{a}}$ & $2027.10$ ($-0.04$) & $5.97$ ($0.09$)\\
         Co$_2$Mn$_{0.9}$Fe$_{0.1}$Ga & $225$ & $4.20$ & $1701.59$ & $2.43$ & $1701.59$ ($0.00$) & $5.15$ ($0.04$)\\
         (Fe$_{0.8}$Mn$_{0.2}$)$_2$MnSn & $225$ & $6.50$ & $1598.79$ & $0.76$ & $1623.15$ ($-0.01$) & $4.04$ ($0.06$) \\
         (Fe$_{0.9}$Mn$_{0.1}$)$_2$CoIn & $225$ & $7.08$ & $1583.29$ & $2.47$ & $1646.07$ ($-0.01$) & $7.27$ ($0.07$)\\
         Fe$_2$ScIn$_{0.8}$Sn$_{0.2}$ & $225$ & $3.98$ & $1473.78$ & $-4.17$ & $1769.44$ ($0.13$) & $5.63$ (0.16)\\
         (Fe$_{0.8}$Mn$_{0.2}$)$_3$Sn & $225$ & $7.15$ & $1472.18$ & $3.30$ & $1787.16$ ($-0.03$) & $5.01$ ($0.04$) \\
         \hline\hline
         \end{tabular}
         \begin{tablenotes}[flushleft]
            \setlength\labelsep{0pt}
            \item[a] Theoretical results obtained with VASP, Ref.~\cite{noky2020giant} 
            \item The full results are tabulated in the Supplemental Material. Contents in the table are the space group (SG), calculated magnetic moment $m$ per formula unit, 
            the $\sigma_{xy}$ and $\alpha_{xy}$ at the Fermi energy, and the maximum value $\sigma_{xy}^{\mathrm{max}}$ and $\alpha_{xy}^{\mathrm{max}}$ obtained
            in an energy range of $[-0.3{:}0.3]$~ eV around the Fermi energy. $\Delta\epsilon$ denotes the energy difference of the maximum values with respect to the 
            Fermi energy.  
         \end{tablenotes}
   \end{threeparttable}
   \end{adjustbox}
\end{table*}

\begin{table*}
   \begin{adjustbox}{max width = 2\columnwidth}
   \begin{threeparttable}
      \centering
         \caption{Identified top 10 promising $L2_1$-type candidates with large $|\alpha_{xy}|$ (at $T=300$~K) at the Fermi energy.}
         \label{table:alp}
         \small
         \renewcommand{\arraystretch}{1.25}
         \begin{tabular}{lccccccc}
         \hline\hline
         Candidates & SG & $m$ & $\sigma_{xy}$ & $\alpha_{xy}$ & $\sigma_{xy}^{\mathrm{max}}$ [$\Delta\epsilon$] &
         $\alpha_{xy}^{\mathrm{max}}$ [$\Delta\epsilon$] \\
                 & &($\mu_{\mathrm{B}}$/f.u.) & (S\,cm$^{-1}$) & (A\,m$^{-1}$K$^{-1}$)  &  (S\,cm$^{-1}$) [(eV)]& (A\,m$^{-1}$K$^{-1}$) [(eV)] \\
         \toprule[0.15mm]
         Rh$_2$Co$_{0.7}$Fe$_{0.3}$In & $225$ &$3.43$ & $394.57$ & $8.57$ & $-1649.50$ ($0.10$) & $9.74$ ($0.03$)\\
         (Co$_{0.7}$Ni$_{0.3}$)$_3$Sn &$225$ & $3.15$ & $-30.66$ & $8.56$ & $-2614.20$ ($0.06$) & $11.48$ ($0.03$) \\
         Rh$_2$NiSi$_{0.7}$P$_{0.3}$ & $225$  & $0.92$ & $-770.92$ & $8.48$ & $-1595.03$ ($0.06$) & $8.79$ ($-0.02$)\\
         (Co$_{0.9}$Ni$_{0.1}$)$_2$FeSn & $225$ &$5.59$ & $-158.04$ & $8.31$ & $-3351.26$ ($0.10$) & $12.65$ ($0.04$) \\
         (Co$_{0.8}$Ni$_{0.2}$)$_2$FeGe & $225$ &$5.31$ & $-1867.45$ & $8.20$ & $-3152.20$ ($0.02$) & $10.90$ ($-0.03$)\\
         Rh$_2$NiSn & $225$ & $1.15$, $1.0^{\mathrm{a}}$ & $-545.48$, $-360^{\mathrm{a}}$ & $7.56$, $8.14^{\mathrm{a}}$ & $-1728.71$ ($0.14$) & $7.56$ ($0.00$)\\
         (Co$_{0.8}$Ni$_{0.2}$)$_2$MnP & $225$ & $5.84$ & $86.84$ & $7.36$ & $-1548.89$ ($0.12$) & $8.41$ ($0.03$)\\
         Rh$_2$Co$_{0.7}$Fe$_{0.3}$Ga & $225$ & $3.44$ & $284.36$ & $6.96$ & $-1844.04$ ($0.10$) & $9.35$ ($0.04$)\\
         Fe$_2$CoIn$_{0.9}$Sn$_{0.1}$ & $225$ & $6.82$ & $802.75$ & $6.81$ & $1815.55$ ($0.27$) & $6.81$ ($0.00$)\\
         Rh$_2$MnGa$_{0.8}$Ge$_{0.2}$ & $225$ & $4.31$ & $1012.05$ & $6.47$ & $2374.07$ ($-0.13$) & $-8.67$ ($-0.22$) \\
         \hline\hline
         \end{tabular}
         \begin{tablenotes}[flushleft]
            \setlength\labelsep{0pt}
            \item[a] Theoretical results obtained with VASP, Ref.~\cite{noky2020giant} 
            \item The caption is the same as in Table~\ref{table:sig}
         \end{tablenotes}
   \end{threeparttable}
   \end{adjustbox}
\end{table*}

\subsection{Identified promising chemically substituted candidates with large AHC and ANC}

Figure~\ref{fig:Fig2} presents the absolute values of $\sigma_{xy}$ and $\alpha_{xy}$ for 1493 candidates, categorized based on the element at the $X$ site and represented by different symbols.
The blue and red areas represent candidates that possess large $|\sigma_{xy}|$ and $|\alpha_{xy}|$, respectively, meeting criteria of $|\sigma_{xy}|=1500$~S\,cm$^{-1}$ 
and $|\alpha_{xy}|=5$~A\,m$^{-1}$K$^{-1}$, which are determined based on the experimentally reported substantial value in Fe~\cite{miyasato2007crossover} and SmCo$_{5}$~\cite{miura2019observation,uchida2021transverse} at room temperature.
From Fig.~\ref{fig:Fig2}, it becomes evident that $\sigma_{xy}$ and $\alpha_{xy}$ can be easily tailored by chemical substitution.  
Out of these candidates, 71 chemically substituted candidates (open symbols) belonging to 4 groups (Co-, Fe-, Rh- and Ru-based compounds) exceed the defined $|\sigma_{xy}|$ or $|\alpha_{xy}|$ criterion, and 3 candidates (green area) 
demonstrate large $|\sigma_{xy}|$ and $|\alpha_{xy}|$ simultaneously. In contrast, only 6 stoichiometric mother compounds (filled symbols) from the Rh- and Fe-based groups meet the discussed criteria. As depicted in Fig.~S2 of the SM, all promising candidates mentioned above belong to the $L2_1$ type with space group number $225$. It should be noted that several promising candidates (labeled in Fig.~\ref{fig:Fig2}) originate from the same stoichiometric mother compound but with slightly different concentrations of substitution at different atomic sites.

\begin{figure*}
   \centering
    \includegraphics[width=0.95\textwidth]{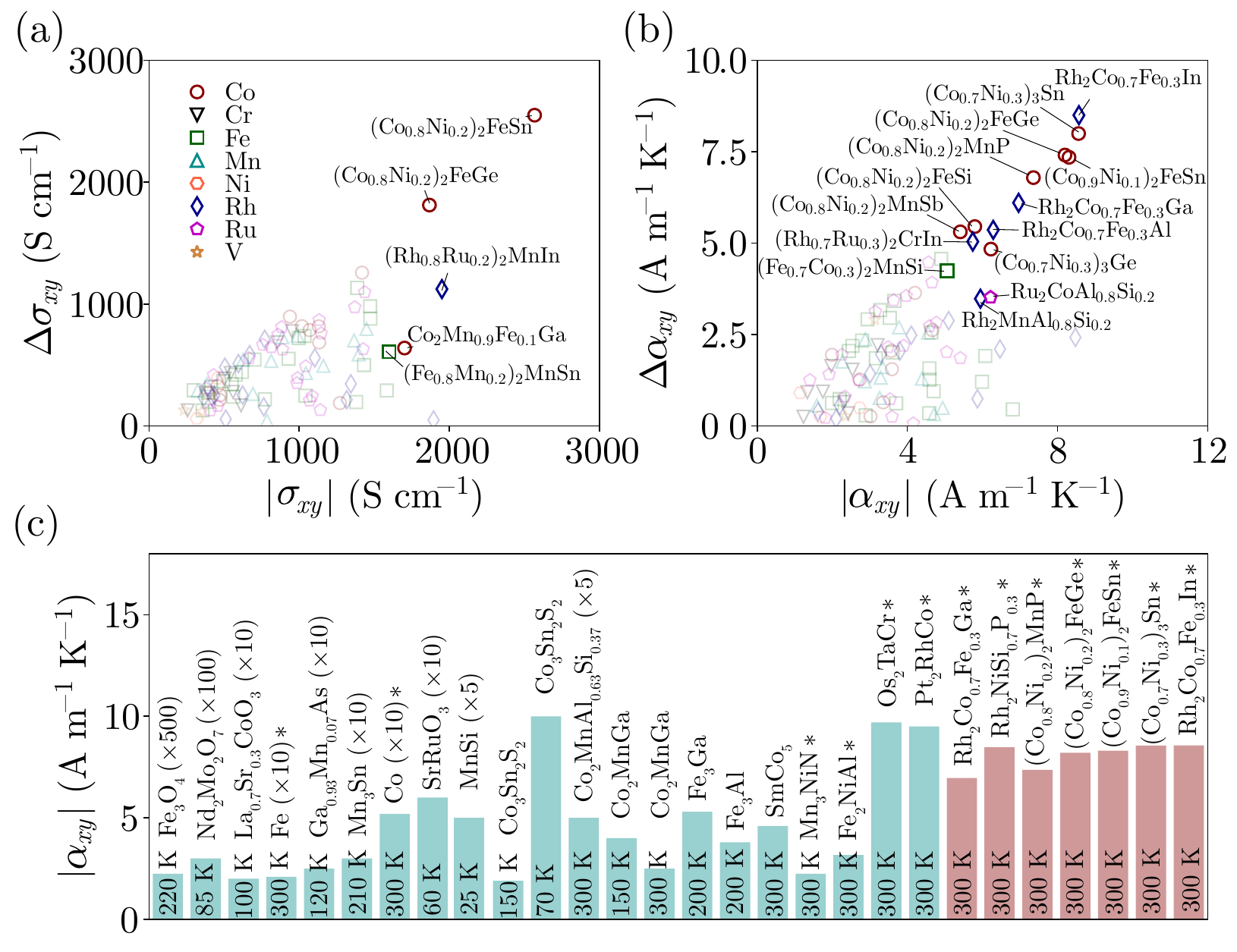}
   \caption{Enhancement of $\sigma_{xy}$ (a) and $\alpha_{xy}$ (b) 
  in the candidates incoparating chemical substitution in comparison with the corresponding stoichiometric mother compounds. Different symbols represent
  the compounds containing the same element at the X site. (c) The calculated $\alpha_{xy}$ for promising candidates (red color) in comparison
  to various magnets (cyan color) taken from previous work~\cite{sakai2018giant,ramos2014anomalous,hanasaki2008anomalous,miyasato2007crossover,weischenberg2013scattering,pu2008mott,ikhlas2017large,li2017anomalous,hirokane2016longitudinal,guin2019zero,yang2020giant,sakuraba2020giant,xu2020anomalous,sakai2020iron,miura2019observation,yang2023intrinsic,tanzim2023giant}. 
  The numbers in parentheses are to scale the original data 
  and asterisks represent the computational results.}
   \label{fig:Fig3}
\end{figure*}

To consider the wide range of potential chemically substituted materials, one candidate with the largest $\sigma_{xy}$ or $\alpha_{xy}$ from each stoichiometric mother compound was selected, and the top 10 candidates are tabulated in Table~\ref{table:sig} and
Table~\ref{table:alp}, respectively. The complete list of materials is provided in the SM. It is evident that most chemically substituted candidates listed in Table~\ref{table:sig} and Table~\ref{table:alp} exhibit significantly enhanced transport quantities compared to the corresponding stoichiometric compounds, with the exception being Rh$_2$MnAl and Rh$_2$NiSn. For stoichiometric mother compounds, we obtained transport properties that are qualitatively similar to those reported by Noky~$et$ $al.$~\cite{noky2020giant}. The differences can be attributed to the use of different pseudopotentials from different DFT codes. Notably, the stoichiometric compound of Co$_2$Mn$_{0.9}$Fe$_{0.1}$Ga, (Fe$_{0.8}$Mn$_{0.2}$)$_2$MnSn, and (Co$_{0.8}$Ni$_{0.2}$)$_2$FeGe, namely Co$_2$MnGa, Fe$_2$MnSn, and Co$_2$FeGe, have been already synthesized by experiments with bulk material or thin films~\cite{sakai2018giant,guin2019anomalous,xu2020anomalous,sumida2020spin,jain2015high,buschow1983magneto}. In particular, (Co$_{0.8}$Ni$_{0.2}$)$_2$FeGe exhibits simultaneously large $\sigma_{xy}$ ($-1867.45$~S\,cm$^{-1}$) and $\alpha_{xy}$ ($8.20$~A\,m$^{-1}$K$^{-1}$). 

We observed the largest and second largest $\sigma_{xy}$ in (Co$_{0.8}$Ni$_{0.2}$)$_2$FeSn ($-2568$~S\,cm$^{-1}$) and (Rh$_{0.8}$Ru$_{0.2}$)$_2$MnIn ($1950$~S\,cm$^{-1}$), respectively. These values significantly surpass the experimentally realized values of $\sim$$1500$~S\,cm$^{-1}$  in Fe~\cite{miyasato2007crossover} and rival those of Co$_{2}$MnGa, which reaches $\sim$$2000$~S\,cm$^{-1}$~\cite{sakai2018giant} at low temperatures. From Table~\ref{table:alp}, Rh$_2$Co$_{0.7}$Fe$_{0.3}$In and (Co$_{0.7}$Ni$_{0.3}$)$_{3}$Sn exhibit the largest $\alpha_{xy}$ values of $\sim$$8.6$~A\,m$^{-1}$K$^{-1}$, followed by Rh$_2$NiSi$_{0.7}$P$_{0.3}$, (Co$_{0.9}$Ni$_{0.1}$)$_2$FeSn, and (Co$_{0.8}$Ni$_{0.2}$)$_2$FeGe with $\alpha_{xy}>8$~A\,m$^{-1}$K$^{-1}$. These magnitudes are substantially larger than those of most typical magnets shown in Fig.~\ref{fig:Fig3} (c) and are comparable to the highest value reported in Co$_3$Sn$_2$S$_2$~\cite{yang2020giant} ($\sim$$10$~A\,m$^{-1}$K$^{-1}$ $70$~K) and in all-$d$ Heusler compounds, such as Os$_2$TaCr ($9.7$~A\,m$^{-1}$K$^{-1}$) and Pt$_2$RhCo ($9.5$~A\,m$^{-1}$K$^{-1}$), at $300$~K~\cite{tanzim2023giant}.
In addition, we found that the values of $\sigma_{xy}$ and $\alpha_{xy}$ in most chemically substituted candidates approach their maximum values (see $\sigma_{xy}^{\mathrm{max}}$ in Table~\ref{table:sig} and $\alpha_{xy}^{\mathrm{max}}$ in Table~\ref{table:alp}) within an energy window of $\sim$0.05~eV. This suggests that the chemical substitution with a concentration of up to $0.3$ can effectively fine-tune and optimize the transport properties of these compounds.

\begin{figure*}
   \centering
   \includegraphics[width=0.95\textwidth]{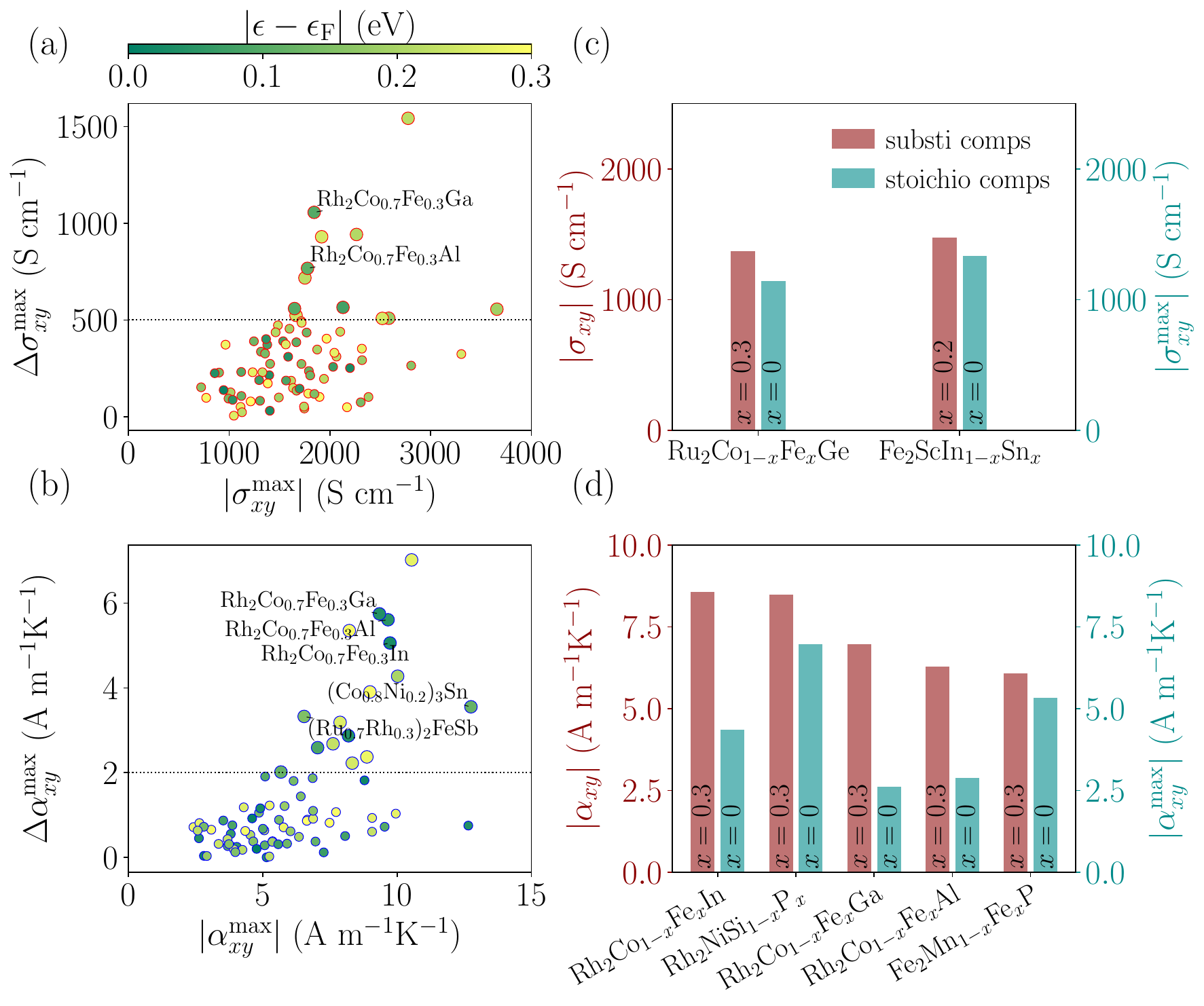}
   \caption{Enhancement of $\sigma_{xy}^{\mathrm{max}}$ (a)
   and $\alpha_{xy}^{\mathrm{max}}$ (b) in chemically substituted candidates in comparison with 
   the corresponding stoichiometric mother compounds. The color bar shows the energy difference of the 
   $\sigma_{xy}^{\mathrm{max}}$ and $\alpha_{xy}^{\mathrm{max}}$ with respect to the Fermi energy.
   Promising chemically substituted candidates (red bars) are identified where their $\sigma_{xy}$ (c) and $\alpha_{xy}$ (d) at the Fermi energy are larger than the corresponding 
   $\sigma_{xy}^{\mathrm{max}}$ and $\alpha_{xy}^{\mathrm{max}}$ in stoichiometric mother compounds (cyan bars) within the energy range of $[-0.3{:}0.3]$~eV around the Fermi energy.}
   \label{fig:Fig4}
\end{figure*}

To quantify the efficiency of chemical substitution, we define an enhancement as $\Delta{T_{xy}}=T_{xy}^{\mathrm{substi}}-T_{xy}^{\mathrm{stoichio}}$, where $T_{xy}^{\mathrm{substi}}$ and $T_{xy}^{\mathrm{stochio}}$ are the absolute value $\sigma_{xy}$ or $\alpha_{xy}$ obtained from the chemically substituted candidate and the corresponding stoichiometric compound, respectively. 
In Fig.~\ref{fig:Fig3}, candidates are displayed with $\Delta\sigma_{xy}\geq500$~S\,cm$^{-1}$ and $|\sigma_{xy}|\geq1500$~S\,cm$^{-1}$ [Fig.~\ref{fig:Fig3}(a)], and with $\Delta\alpha_{xy}\geq3$~A\,m$^{-1}$K$^{-1}$ and $|\alpha_{xy}|\geq5$~A\,m$^{-1}$K$^{-1}$ [Fig.~\ref{fig:Fig3}(b)]. For $\sigma_{xy}$, we have identified 5 candidates shown in Fig.~\ref{fig:Fig3}(a) that meet the defined criteria. In contrast, Fig.~\ref{fig:Fig3}(b) reveals an extensive list of 14 candidates with enhanced $\alpha_{xy}$ values, surpassing the defined criteria. This observation underscores the effective optimization of $\alpha_{xy}$ through chemical substitution. The observed phenomenon is reasonable and can be explained by the entropy density function $s(\epsilon)$ shown in Eq.~(\ref{eq:ANC}) in the integration of BC within the FBZ. At finite temperatures, the entropy density function exhibits nonzero values within an energy window of $|\epsilon-\mu|\lesssim{k_{\mathrm{B}}}T$. Within this window, both the occupied and unoccupied bands near the Fermi energy significantly contribute to the BC. As a result, the BC originating from these bands near the Fermi energy makes $\alpha_{xy}$ amenable to tailoring via chemical substitution. 

In the following, we discuss the possible scenarios that explain the enhancement of $\sigma_{xy}$ and $\alpha_{xy}$ through chemical substitution. 
Initially, the position of the Fermi energy $\epsilon_{\mathrm{F}}$ can be easily tuned through chemical substitution, which introduces carrier (electron or hole), optimizing the transport quantities without changing the electronic structure. 
This straightforward concept aligns with the conventional band-filling approach, which has been widely used for estimating chemical substitution effects. 
In the band-filling approach, the concentration of the substitution is estimated by the shift of $\epsilon_{\mathrm{F}}$ within a rigid band structure of the stoichiometric compound. 
The $\epsilon_{\mathrm{F}}$ in the chemically substituted candidates is determined by integrating the density of states up to the energy level equivalent to the number of valence electrons in the system.

However, it is essential to emphasize that the applicability of the band-filling approach is limited to specific compounds. 
As an illustration, let us consider three cases where Ni substitutes Co in Co$_2$MnGa and Co$_2$FeSn, and Fe substitutes Co in Rh$_2$CoIn. We observed a remarkably similar trend in the energy-dependent $\sigma_{xy}$ for (Co$_{0.9}$Ni$_{0.1}$)$_2$MnGa shown in Fig.~S3(a) of the SM, when employing both the VCA and band-filling approaches. The calculated values of $734.67$ and $765.04$~S\,cm$^{-1}$ obtained from VCA and band-filling approach, respectively, indicate the similarity of both approaches for Co$_2$MnGa. However, a notable discrepancy arises when evaluating the $\sigma_{xy}$ of (Co$_{0.9}$Ni$_{0.1}$)$_2$FeSn [refer to Fig.~S3(b)].In this case, the results obtained through VCA and band filling deviate considerably, yielding $\sigma_{xy}$ values of $-158.04$ and $-51.62$~S\,cm$^{-1}$, respectively. 

This discrepancy is further emphasized in the case of Rh$_2$Co$_{0.7}$Fe$_{0.3}$In. The energy-dependent $\sigma_{xy}$ curves diverge dramatically between the two methods, with the band-filling approach yielding $\sigma_{xy}$ value of $112.58$~S\,cm$^{-1}$ and the VCA approach showing $394.57$~S\,cm$^{-1}$ at the Fermi energy, as illustrated in Fig.~S3(c). This discrepancy can be attributed to substantial changes in the electronic structure brought by chemical substitution. Such cases, particularly where extensive changes in the electronic structure occur, necessitate the use of VCA, as illustrated in Fig.~S3(b) and (c), where band filling is inadequate. To assess the applicability of the band-filling approach, we examined the relationship between the total number of electrons at $X$ and $Y$ sites, denoted as $N_{\mathrm{v}}^{XY}$, and the transport quantities. Our findings suggest that the band-filling approach may not be suitable for most candidates with substitution at the $X$ or $Y$ site, especially for those with $N_{\mathrm{v}}^{XY}$$\sim$~$26.5$. Furthermore, the significant $\sigma_{xy}$ values for (Co$_{0.8}$Ni$_{0.2}$)Fe$Z$ ($Z=$~Sn, Ge) predicted by the VCA (refer to Table~\ref{table:sig}) at $N_{\mathrm{v}}^{XY}=26.4$ are unattainable via the band-filling approach. This discrepancy further underscores the importance of accurately calculating anomalous transport quantities using the VCA approach. Additionally, we observed that predicting $\alpha_{xy}$ using the band-filling approach is less reliable than predicting $\sigma_{xy}$. For further details, please refer to the SM. 

The change in the electronic structure plays a central role in the significant enhancement of $\sigma_{xy}$ and $\alpha_{xy}$ observed in this study. The change can be effectively quantified by analyzing the difference in the maximum transport quantities, denoted as $\Delta{T_{xy}^{\mathrm{max}}}=T_{\mathrm{substi}}^{\mathrm{max}}-T_{\mathrm{stoichio}}^{\mathrm{max}}$, between the chemically substituted candidates and their respective stoichiometric compounds. 
Here, $T_{\mathrm{substi}}^{\mathrm{max}}$ ($T_{\mathrm{stoichio}}^{\mathrm{max}}$) represents the maximum absolute value of either $\sigma_{xy}$ or $\alpha_{xy}$ within an energy range of $[-0.3{:}0.3]$~eV around the $\epsilon_{\mathrm{F}}$ for the chemically substituted (stoichiometric) compounds. 
A positive $\Delta{T_{xy}^{\mathrm{max}}}$ denotes the noteworthy enhancement in the maximum transport quantities for chemically substituted candidates compared to their corresponding stoichiometric counterparts. 
This enhancement stems from the distinct distribution of BC in the FBZ, which, in turn, is influenced by the altered electronic structure.

As illustrated in Fig.~{\ref{fig:Fig4}}(a) and (b), among the $104$ candidates, 79 and 78 exhibited a positive value of $\Delta{\sigma_{xy}^{\mathrm{max}}}$ and 
$\Delta\alpha_{xy}^{\mathrm{max}}$, respectively. This observation highlights the importance of going beyond the band-filling approach to thoroughly understand the effect of chemical substitution. Additionally, Fig.~\ref{fig:Fig4}(a) and (b) display the composition of the candidates, focusing on those with $\Delta{\sigma_{xy}^{\mathrm{max}}}\geq{600}$~S\,cm$^{-1}$ and $\Delta{\alpha_{xy}^{\mathrm{max}}}\geq{3}$~A\,m$^{-1}$K$^{-1}$ within an energy window of $0.15$~eV around $\epsilon_{\mathrm{F}}$.

Of particular interest, this study identified a number of promising candidates in which the values of $\sigma_{xy}$ ($\alpha_{xy}$) for the candidates incorporating chemical substitution [indicated by red bars in Fig.~\ref{fig:Fig4}(c) and (d)] at the Fermi energy exceeded the corresponding $\sigma_{xy}^{\mathrm{max}}$ ($\alpha_{xy}^{\mathrm{max}}$) values of their stoichiometric mother compounds 
[depicted by cyan bars in Fig.~\ref{fig:Fig4}(c) and (d)] within the energy range of $[-0.3{:}0.3]$~eV surrounding the Fermi energy. Notably, for instance, in the case of $0.3$ Fe-substituted Rh$_2$Co$Z$ ($Z=$~Al, Ga, and In) shown in Fig.~\ref{fig:Fig4}(d), the $|\alpha_{xy}|$ values were found to be twice as large as the $|\alpha_{xy}^{\mathrm{max}}|$ in the stoichiometric mother compounds. 
This discrepancy implies a substantial alteration in the band structure upon Fe-substituted Rh$_2$Co$Z$, resulting in a dramatic enhancement of BC.

\subsection{Significant change of electronic structure through chemical substitution}

\begin{figure*}
   \centering
   \includegraphics[width=0.95\textwidth]{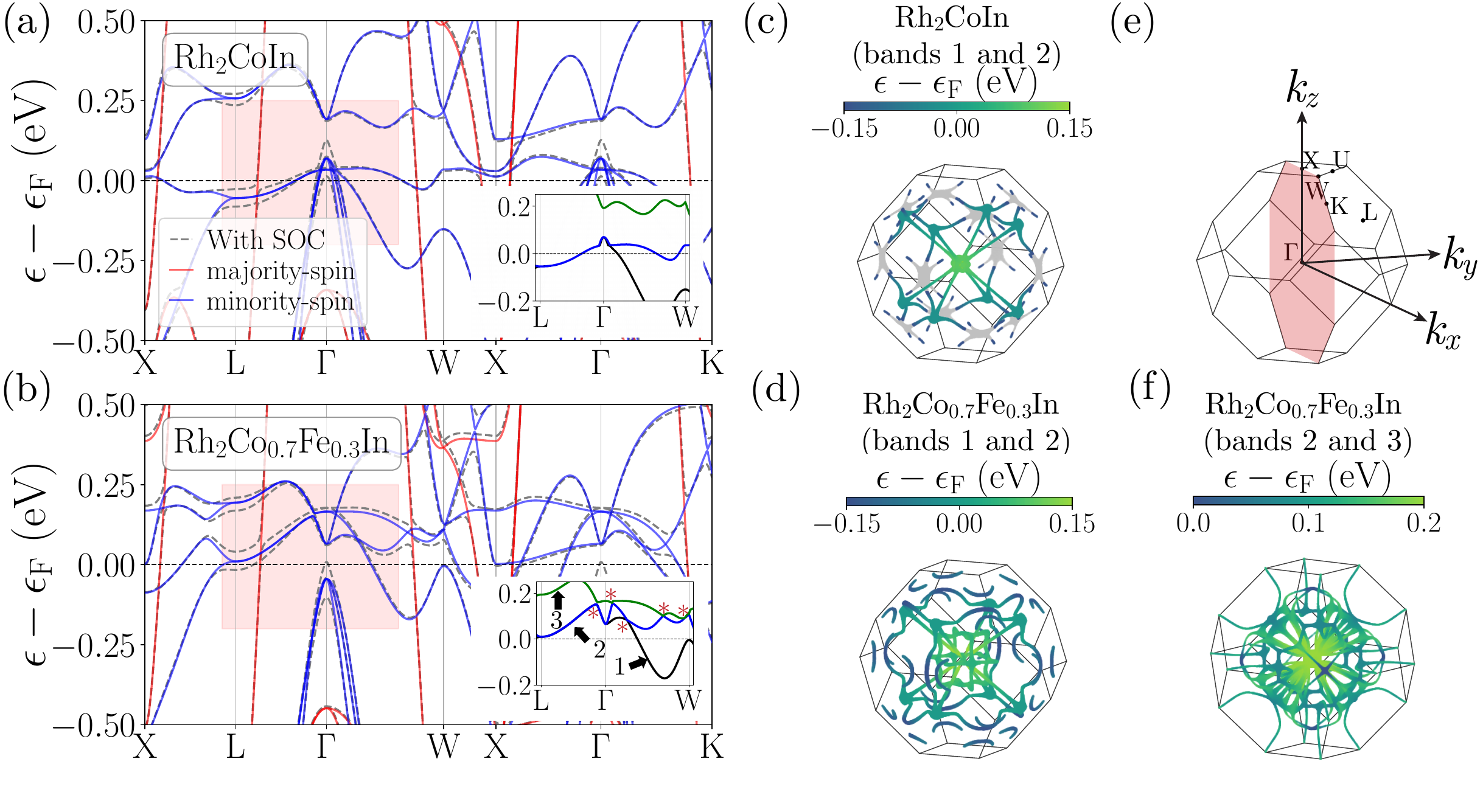}
   \caption{Band structure of Rh$_2$CoIn (a) and Rh$_2$Co$_{0.7}$Fe$_{0.3}$In (b), respectively. The red and blue curves 
  represent the majority- and minority-spin bands, respectively. Dashed gray cures are the 
  bands computed with SOC. The inset shows the minority-spin bands in the shaded area that
  form nodal line networks. Asterisks show the position of the band crossing point in the nodal line. Nodal line network of Rh$_2$CoIn formed by bands 1 and 2 (c), and  Rh$_2$Co$_{0.7}$Fe$_{0.3}$In formed by bands 1
  and 2 (d) and bands 2 and 3 (f). The color bar represents the nodal line energy window around the Fermi energy, with energies outside this window shown in gray. Note that bands 2 and 3
  in Rh$_2$CoIn are gapped out. (e) The first Brillouin with high symmetry points and the $k_y=0$ plane are shown with the shaded area.} 
   \label{fig:Fig5}
\end{figure*}

\begin{figure*}
   \centering
   \includegraphics[width=0.95\textwidth]{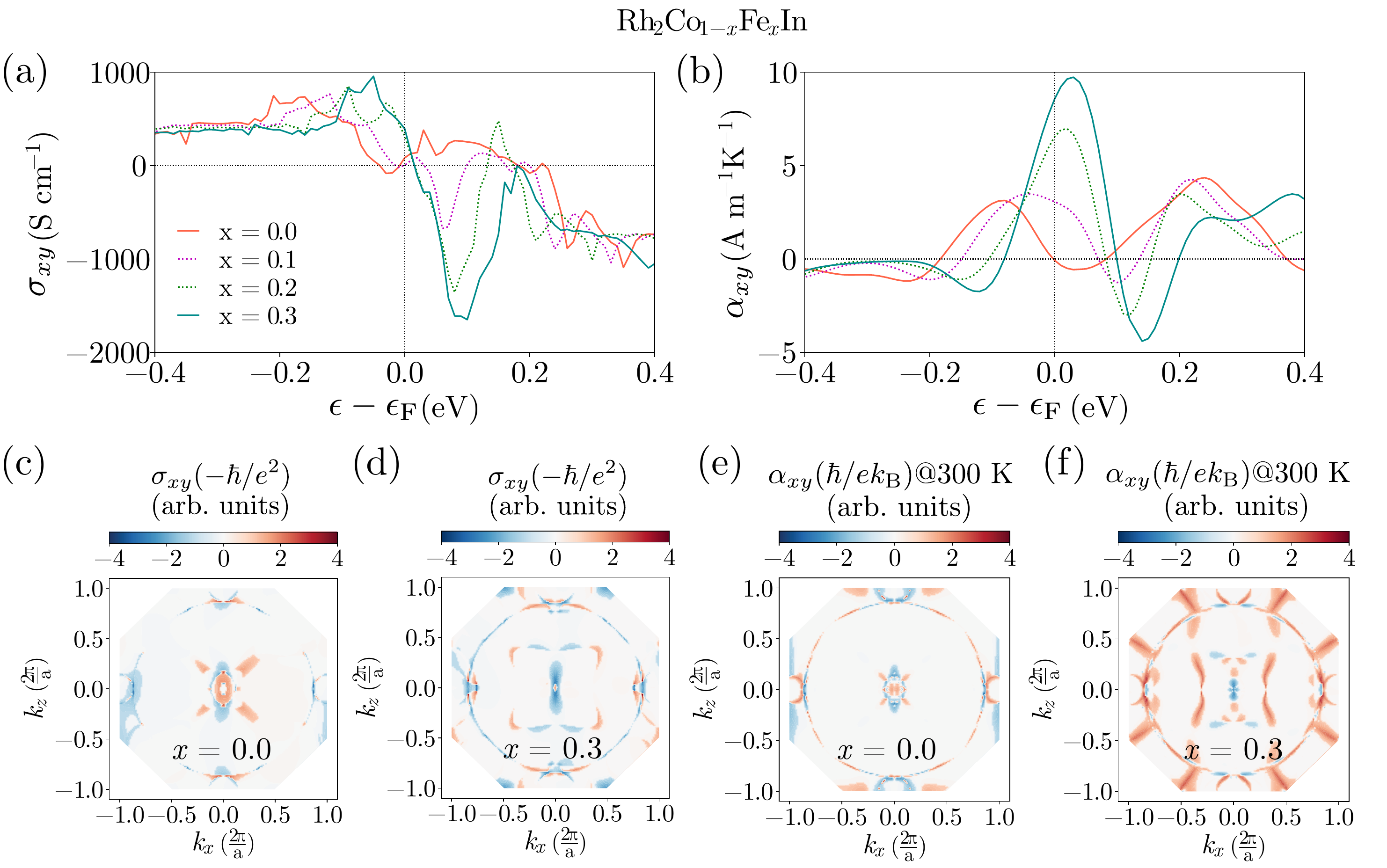}
   \caption{Energy-dependent $\sigma_{xy}$ (a) and $\alpha_{xy}$ (b) of Rh$_2$Co$_{1-x}$Fe$_x$In with $x=0$, $0.1$, $0.2$, $0.3$, respectively.
  $\bm{k}$-decomposed $\sigma_{xy}$ (c) and (d) and $\alpha_{xy}$ (e) and (f) at $k_y=0$ plane for Rh$_2$Co$_{1-x}$Fe$_x$In
 with $x=0$ and $x=0.3$, respectively. Note that the color scale for $\sigma_{xy}$ and
  $\alpha_{xy}$ is the same.}
   \label{fig:Fig6}
\end{figure*} 

In this section, we provide a detailed discussion of three highly promising candidates that belong to different groups of $X$ elements, each exhibiting substantial $\alpha_{xy}$ values. These candidates are Rh$_2$Co$_{0.7}$Fe$_{0.3}$In, Fe$_2$Mn$_{0.7}$Fe$_{0.3}$P, and (Co$_{0.7}$Ni$_{0.3}$)$_3$Sn. All of these candidates share the $L2_1$-type crystal structure~\cite{L21}, hosting three mirror planes at $x=0$, $y=0$, and $z=0$. Previous studies~\cite{noky2020giant,noky2019large,shukla2021anomalous} have reported the presence of three symmetry-protected nodal lines (NL) were observed at $k_x=0$, $k_y=0$, and $k_z=0$ planes without including the SOC. Among these, the NL at $k_z=0$ plane remains protected, while others at $k_x=0$ and $k_y=0$ planes become gapped when SOC is introduced, aligning the magnetization direction along [001] direction. The small band gap surrounding the former NL induces a strong BC. Consequently, the significant $\sigma_{xy}$ or $\alpha_{xy}$ values are observed when the energy of the gapped NL is distributed around $\epsilon_{\mathrm{F}}$. 

In addition to the previously discussed NLs, we have identified more complex NL networks near $\epsilon_{\mathrm{F}}$ resulting from chemical substitution. 
Specifically, in the case of Rh$_2$Co$_{0.7}$Fe$_{0.3}$In, the complex NL induced by chemical substitution results in an almost twofold increase in the $\alpha_{xy}$ ($8.57$~A\,m$^{-1}$K$^{-1}$) at the Fermi energy when compared to the peak value ($4.36$~A\,m$^{-1}$K$^{-1}$) of the stoichiometric compound Rh$_2$CoIn.

\subsubsection{Nodal line network creation in Rh$_2$CoIn through chemical substitution}

Before delving into the discussion of the band structures, we direct our attention to the magnetic moment of both the chemically substituted and stoichiometric compounds of Rh$_2$CoIn. 
Within Rh$_2$CoIn, the computed total magnetic moment amounts to $3.02$ $\mu_{\mathrm{B}}$ per formula unit (f.u.), accompanied by local moments of $2.08$ and $0.52$ $\mu_{\mathrm{B}}$ at the corresponding Co and Rh sites, respectively. Our total magnetic moment is in good agreement with that calculated by Noky~$et$ $al.$ ($3.0$~$\mu_{\mathrm{B}}$/f.u.)~\cite{noky2020giant}, alongside local moments of 2.02 and 0.52~$\mu_{\mathrm{B}}$ at the Co and Rh sites, respectively.
The total moment increases to $3.43$~$\mu_{\mathrm{B}}$/f.u. with $0.3$ Fe substituting to the Co site, primarily resulting from the enhanced local moment of $2.48$~$\mu_{\mathrm{B}}$ at pseudo atom Co/Fe site. 

Figure~\ref{fig:Fig5} (a) and (b) illustrate the band structure of Rh$_2$CoIn and  Rh$_2$Co$_{0.7}$Fe$_{0.3}$In, respectively, where the solid curves represent the majority (red) and minority (blue) spin-polarized bands, while the dashed gray curves depict the bands computed with SOC. 
Notably, a distinct contrast in the minority-spin bands between the chemically substituted candidates and the stoichiometric mother compounds is evident within the shaded area of Fig.~\ref{fig:Fig5}(a) and (b), highlighting the following significant changes. First, in Rh$_2$CoIn, the two minority bands [as shown by the black and blue bands in the inset of Fig.~\ref{fig:Fig5}(a)] crossing $\epsilon_{\mathrm{F}}$ are shifted upwards upon Fe substitution. This is a reasonable consequence of $\epsilon_{\mathrm{F}}$ shifting to lower energy due to introducing hole carrier. However, in Rh$_2$Co$_{0.7}$Fe$_{0.3}$In, the corresponding bands, labeled as 1 and 2 in the inset of Fig.~\ref{fig:Fig5}(b), undergo dramatic modifications, forming a
band crossing point along the $\Gamma$--$\mathrm{W}$ high-symmetry path. Second, bands 2 and 3 in Rh$_2$Co$_{0.7}$Fe$_{0.3}$In, as depicted in the inset of Fig.~\ref{fig:Fig5}(b) intersect and give rise to several 
band crossing points along the $\mathrm{L}$--$\Gamma$--$\mathrm{W}$ high-symmetry path. It is worth noting that bands 2 and 3 in the stoichiometric mother compound exhibit a gap [see the inset of Fig.~\ref{fig:Fig5}(a)].

The NL networks formed by the bands 1 and 2 for Rh$_2$CoIn and Rh$_2$Co$_{0.7}$Fe$_{0.3}$In are visualized in Fig.~\ref{fig:Fig5}(c) and (d), respectively, with the energy difference from $\epsilon_{\mathrm{F}}$ indicated by the color bar. The NL networks are analyzed without considering the SOC to understand the topological properties. The degeneracy along $\Gamma$--$\mathrm{L}$ in the NL network is observed for both cases. However, a distinct feature emerges in the case with chemical substitution where three nodal loops are formed at $k_x=0$, $k_y=0$, and $k_z=0$ mirror planes. Intriguingly, a more complex NL network, formed by bands 2 and 3 is exclusively identified in the chemically substituted candidate, as shown in Fig.~\ref{fig:Fig5}(f).
The NL network's energy resides above the Fermi energy within a range of $0.2$~eV, suggesting its contribution solely to $\alpha_{xy}$, as further elaborated below.

Next, we discuss the underlying factors behind the notable $\alpha_{xy}$ observed in Rh$_2$Co$_{0.7}$Fe$_{0.3}$In. To comprehensively understand the relationship between $\sigma_{xy}$ and $\alpha_{xy}$, $\alpha_{xy}$ can be 
formulated as,
\begin{align}
   \alpha_{xy} =-\frac{k_{\mathrm{B}}}{e}\int{}d\epsilon{}s(\epsilon)\frac{\partial{\sigma_{xy}(\epsilon,T=0)}}{\partial{\epsilon}}.
   \label{eq:ANC2}
\end{align}
As Eq.~(\ref{eq:ANC2}) illustrates, the magnitude of $\alpha_{xy}$ is directly proportional to the energy derivative of $\sigma_{xy}$ weighted by the entropy density function $s(\epsilon)$. Given $s(\epsilon)$ is like a smeared delta function around
$\epsilon_{\mathrm{F}}$~\cite{xu2020anomalous}, $\alpha_{xy}$ can be roughly estimated from $\partial{\sigma_{xy}}/\partial{\epsilon}$ around $\epsilon=\epsilon_{\mathrm{F}}$.

Figure~\ref{fig:Fig6} (a) and (b) present the energy dependence of $\sigma_{xy}$ and $\alpha_{xy}$ within Rh$_2$Co$_{1-x}$Fe$_x$In, with $x$ ranging from $0.0$ to $0.3$ in increments of $0.1$. 
In Fig.~\ref{fig:Fig6}(a), a distinct feature of Fe substitution is observed: the peak value of Rh$_2$CoIn, initially near $\sim$$-0.2$~eV, gradually shifts to higher energy with increasing $x$.  This shift is accompanied by the emergence of a negative slope around the Fermi energy as $x$ increases. Eventually, in Rh$_2$Co$_{0.7}$Fe$_{0.3}$In, a substantial peak value of $-1649.50$~S\,cm$^{-1}$ is achieved at $0.1$~eV above the Fermi energy. 
In contrast, the stoichiometric mother compound lacks this negative peak.
The consequential negative energy-dependent slope in $\sigma_{xy}$ leads to a giant $\alpha_{xy}$ of $8.57$~A~m$^{-1}$K$^{-1}$ at $300$~K shown in Fig.~\ref{fig:Fig6}(b).

We investigated the $\bm{k}$-decomposed $\sigma_{xy}$ and $\alpha_{xy}$ at $k_y=0$ plane, where the NL becomes gapped in the presence of SOC. This analysis provides deeper insights into the substantial $\alpha_{xy}$ observed in Rh$_2$Co$_{0.7}$Fe$_{0.3}$In. When examining the $\sigma_{xy}$ for both compounds in Fig.~\ref{fig:Fig6}(c) and (d) and $\alpha_{xy}$ for Rh$_2$CoIn in Fig.~\ref{fig:Fig6}(e), a distinct lack of large values in the plane becomes evident. In contrast, Fig.~\ref{fig:Fig6}(f) reveals a noticeable positive $\bm{k}$-decomposed pattern with significant $\alpha_{xy}$ values for Rh$_2$Co$_{0.7}$Fe$_{0.3}$In.
This pattern is consistent with the NL loops around $\Gamma$ and $\mathrm{W}$ points [see the position in Fig.~\ref{fig:Fig5}(e)], originating from the gapped NL networks shown in Fig.~\ref{fig:Fig5} (d) and (f) exclusively to the chemically substituted materials.
It is worth noting that the energy of the mentioned NL networks is mainly located above $\epsilon_{\mathrm{F}}$, rendering them irrelevant to the contribution of $\sigma_{xy}$.

\subsubsection{Nodal line network modification through chemical substitution}

\begin{figure*}
   \centering
   \includegraphics[width=\textwidth]{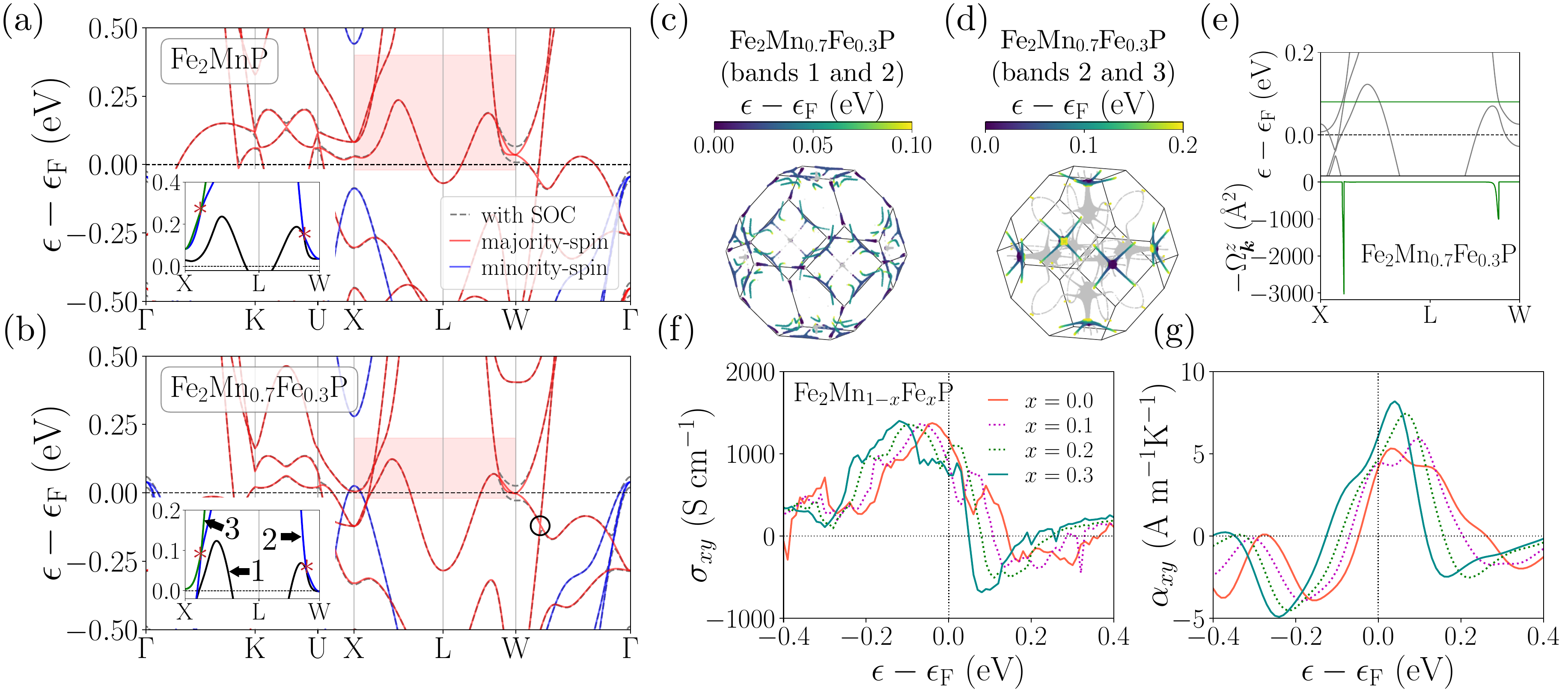}
   \caption{Band structure of Fe$_2$MnP (a) and  Fe$_2$Mn$_{0.7}$Fe$_{0.3}$P (b), respectively.
  The red and blue curves represent the majority- and minority-spin bands, respectively. Dashed gray cures are the 
  bands computed with SOC. The inset shows the majority-spin bands in the shaded area that
  form nodal line networks. Asterisks show the position of the band crossing point in the nodal line.  
  Nodal line network of Fe$_2$Mn$_{0.7}$Fe$_{0.3}$P formed by bands 1 and 2 (c),
  and bands 2 and 3 (d), respectively. The color bar represents the nodal line energy window around the Fermi energy, with energies outside this window shown in gray. (e) Berry curvature $\Omega_{\bm{k}}^z$ evaluated at $0.08$~eV (green line) above the Fermi energy along the high-symmetry path. Energy-dependent $\sigma_{xy}$ (f) and $\alpha_{xy}$ (g)
  for Fe$_2$Mn$_{1-x}$Fe$_x$P with $x=0$, $0.1$, $0.2$, $0.3$, respectively.}
   \label{fig:Fig7}
\end{figure*} 

\begin{figure*}
   \centering
   \includegraphics[width=0.95\textwidth]{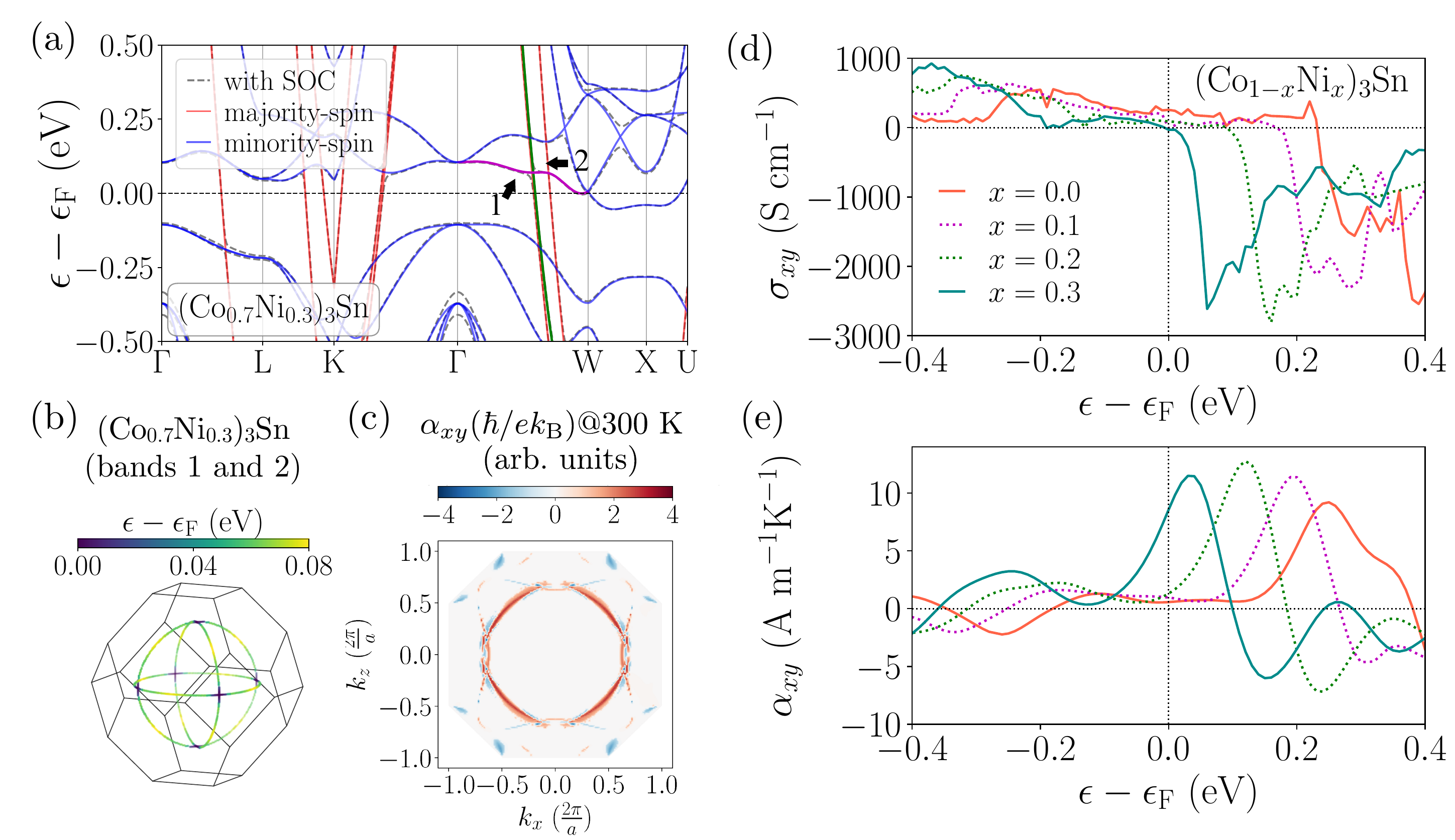}
   \caption{(a) Band structure of (Co$_{0.7}$Ni$_{0.3}$)$_3$Sn. The red and blue curves represent the majority- and minority-spin bands, respectively. Dashed gray cures are the 
  bands computed with SOC. The minority-spin band labeled 1 and the majority-spin band labeled 2 formed the nodal line. (b) Nodal line network
of (Co$_{0.7}$Ni$_{0.3}$)$_3$Sn formed by bands 1 and 2. The color bar represents the nodal line energy
window around the Fermi energy. (c) $\bm{k}$-decomposed $\alpha_{xy}$ of at $k_y=0$ plane for (Co$_{0.7}$Ni$_{0.3}$)$_3$Sn. Energy-dependent $\sigma_{xy}$ (d) and $\alpha_{xy}$ (e) of (Co$_{1-x}$Ni$_x$)$_3$Sn with $x=0$, $0.1$, $0.2$, $0.3$, respectively.}
   \label{fig:Fig8}
\end{figure*}

The creation of NL networks through chemical substitution, as observed in the Rh$_2$Co$_{1-x}$Fe$_x$In, is not a common occurrence. 
In most general cases, the Fermi energy can be effectively tuned via chemical substitution. Simultaneously, the band structure undergoes slight modifications due to chemical substitution, impacting the energy-dependent behavior of BC and subsequently leading to alterations in the transport quantities. 
Here, we take the Fe$_2$Mn$_{0.7}$Fe$_{0.3}$P and (Co$_{0.7}$Ni$_{0.3}$)$_3$Sn as examples, to illustrate the enhancement of $\alpha_{xy}$ through the shift of the Fermi energy accompanied by the band structure modifications via chemical substitution. 

The ferromagnetic state was observed in Fe$_2$MnP, where the total magnetic moment is $3.97$~$\mu_{\mathrm{B}}$/f.u., with local moments of $2.55$ and $0.69$~$\mu_{\mathrm{B}}$ at Mn and Fe sites, respectively. 
The observed moments are consistent with reported values ($\mu_{\mathrm{tot}}=4$~$\mu_{\mathrm{B}}$/f.u., $\mu_{\mathrm{Mn}}=2.6$~$\mu_{\mathrm{B}}$, $\mu_{\mathrm{Fe}}=0.7$~$\mu_{\mathrm{B}}$, Ref.~\cite{noky2019large}). 
A total moment of $4.30$~$\mu_{\mathrm{B}}$/f.u. was obtained for Fe$_2$Mn$_{0.7}$Fe$_{0.3}$P, with an increase of $\sim$$0.1$~$\mu_{\mathrm{B}}$ at each Mn/Fe and Fe site. In contrast, the total moment of $4.16$~$\mu_{B}$/f.u. ($4.04$~$\mu_{\mathrm{B}}$ in Ref.~\cite{noky2020giant}) in Co$_3$Sn, which decreases to $3.15$~$\mu_{\mathrm{B}}$/f.u. when 0.3 Ni was substituted at the Co site.

The band structures of Fe$_2$MnP and Fe$_2$Mn$_{0.7}$Fe$_{0.3}$P are presented in Fig.~\ref{fig:Fig7} (a), and (b), respectively. It is evident that the overall band structures for these two cases exhibit notable similarities, with only minor modifications such as the location of $\epsilon_{\mathrm{F}}$ and the band dispersion along specific high-symmetry paths. 
Interestingly, these subtle adjustments lead to the enhancement of $\alpha_{xy}$ in the chemically substituted candidates, a point we will elaborate upon. 
In both compounds, the intriguing topological properties stem from the majority-spin bands, highlighted in the shaded areas of Fig.~\ref{fig:Fig7}(a) and (b), respectively. Given the similarity of NL networks in both chemically substituted and mother compounds, here we focus on the Fe$_2$Mn$_{0.7}$Fe$_{0.3}$P compound. 

We have identified three distinct types of NL networks near the Fermi energy. The first, formed
by two majority bands [indicated by the black circle in Fig.~\ref{fig:Fig7}(b)], is shown in Fig.~S5 of the SM. The NL networks are located at the three mirror planes of $k_x=0$, $k_y=0$, and $k_z=0$, respectively, with their energies mainly distributed below the Fermi energy. 
This specific NL has been extensively discussed by Noky $et$ $al.$~\cite{noky2019large} and serves as the main reason for the large $\sigma_{xy}$ for both compounds. As depicted in Fig.~\ref{fig:Fig7}(f), the $\sigma_{xy}^{\mathrm{max}}$, with the values of $\sim$$1400$~S\,cm$^{-1}$ for both compounds mainly originates from the BC induced by this gapped NL.

The second and third NL networks, depicted in Fig.~\ref{fig:Fig7}(c) and (d), are formed by bands 1 and 2, and bands 2 and 3 [refer to the inset of Fig.~\ref{fig:Fig7}(b)], respectively. These NL networks are less commonly discussed, primarily due to their NL energies predominantly residing mainly above $\epsilon_{\mathrm{F}}$, which makes a minor contribution to $\sigma_{xy}$.
Nevertheless, we observed subtle changes in the band dispersion of these bands upon chemical substitution. However, a discernible difference appears in the energy of the band crossing points along $\mathrm{X}$--$\mathrm{L}$--$\mathrm{W}$ high-symmetry path between the chemically substituted and stoichiometric compounds. This energy difference is pivotal in explaining the significant $\alpha_{xy}$ observed in Fe$_2$Mn$_{0.7}$Fe$_{0.3}$P. 

In Fe$_2$MnP,
the two band crossing points shown in the inset of Fig.~\ref{fig:Fig7}(a), formed by bands 1 and 2, and bands 2 and 3, are situated considerably away from $\epsilon_{\mathrm{F}}$, at $0.17$ and $0.3$~eV above $\epsilon_{\mathrm{F}}$, respectively. Through substituting Fe at the Mn site, both band crossing points shown in the inset of Fig.~\ref{fig:Fig7}(b) shift towards the Fermi energy. Importantly, these two points nearly converge to an energy level of $\epsilon-\epsilon_{\mathrm{F}}$$\sim$$0.08$~eV. As a consequence, two significant BCs are simultaneously induced at $0.08$~eV above $\epsilon_{\mathrm{F}}$ along the $\mathrm{X}$--$\mathrm{L}$--$\mathrm{W}$ high-symmetry path by the aforementioned points in the presence of SOC, as demonstrated in Fig.~\ref{fig:Fig7}(e). These substantial BCs result in a steep negative slope in the energy-dependent $\sigma_{xy}$ within $0.08$~eV above $\epsilon_{\mathrm{F}}$, as depicted in Fig.~\ref{fig:Fig7}(f). This slope gives rise to a considerable $\alpha_{xy}$ value of $6.08$~A\,m$^{-1}$K$^{-1}$ at $\epsilon_{\mathrm{F}}$ in Fe$_2$Mn$_{0.7}$Fe$_{0.3}$P, as illustrated in Fig.~\ref{fig:Fig7}(g).

Importantly, it is worth noting that this $\alpha_{xy}$ is not the peak, as the maximum value of $8.19$~A\,m$^{-1}$K$^{-1}$ is found merely $0.04$~eV above $\epsilon_{\mathrm{F}}$. In contrast, even the $\alpha_{xy}^{\mathrm{max}}$ in the stoichiometric compound is $5.32$~A\,m$^{-1}$K$^{-1}$ [at $0.03$~eV above $\epsilon_{\mathrm{F}}$ in Fig.~\ref{fig:Fig7}(g)], highlighting the significant impact of minor NL network modifications stemming from chemical substitution, which subsequently induce considerable $\alpha_{xy}$ values. For further details on the relatively modest $\alpha_{xy}$ in Fe$_2$MnP, please refer to the SM.

Finally, Ni-substituted Co$_3$Sn is discussed. Our analysis reveals that the bands forming the NL network arise from the different spin states. We note that the NL formed by different spin states occupies a narrow energy range. Through chemical substitution, we strategically shift the Fermi energy to align with the NL energy, consequently resulting in the generation of substantial $\sigma_{xy}$ and $\alpha_{xy}$ values. 
Intriguingly, we observed the common feature in (Co$_{1-x}$Ni$_{x}$)$_2$FeSn [refer to the energy-dependent anomalous transport properties in Fig.~S7 of the SM], and among various Ni-substituted Co$_2XZ$ systems ($Y=$~Co, Fe; $Z=$~Sn, Ge) with substantial $\sigma_{xy}$ or $\alpha_{xy}$ values shown in Fig.~\ref{fig:Fig2}. 
Illustrating this with (Co$_{0.7}$Ni$_{0.3}$)$_3$Sn, we observed in Fig.~\ref{fig:Fig8} (a) the intersection of minority- and majority-spin bands, labeled as 1 and 2 respectively, along the $\Gamma$--$\mathrm{W}$ high-symmetry path. These intersecting bands give rise to three NLs, as depicted in Fig.~\ref{fig:Fig8}(b), situated at the previously mentioned mirror planes.
It is important to note that the energy of these NLs resides above $\epsilon_{\mathrm{F}}$, encompassed within a narrow energy window of $0.08$~eV. 

The presence of a robust BC becomes apparent within this energy window, originating from the gapped NLs with SOC. Consequently, we observe a distinct negative peak value of $-2614.20$~S\,cm$^{-1}$ in $\sigma_{xy}$ for (Co$_{0.7}$Ni$_{0.3}$)$_3$Sn,
located $0.06$~eV above $\epsilon_{\mathrm{F}}$, as evident in Fig.~\ref{fig:Fig8}(d). A remarkable $\alpha_{xy}$ of $8.56$~A\,m$^{-1}$K$^{-1}$ [see Fig.~\ref{fig:Fig8}(e)] was obtained, with clear $\bm{k}$-decomposed $\alpha_{xy}$ distributed along the gapped NLs, as illustrated in Fig.~\ref{fig:Fig8}(c). Worth mentioning is the energy in the corresponding NLs of the stoichiometric mother compounds, located $\sim$$0.3$~eV above $\epsilon_{\mathrm{F}}$, resulting in a negligible contribution to both $\sigma_{xy}$ and $\alpha_{xy}$.

\section{Summary}

In summary, our study involves an extensive high-throughput calculation of 1493 magnetic cubic Heusler compounds, incorporating chemical substitution through first-principles calculations, aimed at identifying promising candidates with enhanced transport properties. These candidates are categorized based on the element at the $X$ site. Notably, compounds based on Co and Rh demonstrate remarkable potential for optimization of the transport property. 
Specifically, the compound (Co$_{0.8}$Ni$_{0.2}$)$_2$FeSn exhibits exceptional perfomance, with both giant $\sigma_{xy}$ and $\alpha_{xy}$ ($T=300$~K) values of $-2567$~S\,cm$^{-1}$ and $8.27$~A\,m$^{-1}$K$^{-1}$, respectively. These values surpass both experimental and theoretical results reported for other typical magnets~\cite{miyasato2007crossover,tanzim2023giant,noky2020giant,tung2013high}.  
Additionally, we highlight that 5 candidates derived from distinct stoichiometric mother compounds exhibit $\alpha_{xy}$ values exceeding $8$~A\,m$^{-1}$K$^{-1}$. The magnitude of these values is comparable to those reported for Co$_3$Sn$_2$S$_2$~\cite{yang2020giant} and several recently explored all-$d$ Heusler compounds~\cite{tanzim2023giant}.

Furthermore, we emphasize the pivotal role of evaluating the chemical substition effects using VCA, as opposed to relying solely on the band-filling approach with stoichiometric mother compounds. The approach we employed not only takes into account the shift in Fermi energy but also counts the significant modification in the band structure through chemical substitution. 
As a result, we have identified a number of promising candidates, especially in the case of Rh$_2$Co$_{0.7}$Fe$_{0.3}Z$ ($Z=$~Al, Ga, In), where the $\alpha_{xy}$ exceed the maximum observed in the stoichiometric mother compounds. 
Through a detailed analysis of the band structure, we revealed that the creation or modification of nodal lines via chemical substitution leads to a strong BC slightly above the Fermi energy and consequently yields substantial $\alpha_{xy}$. 
Our work provides valuable insights into the optimization of transport properties by chemical substitution, thereby paving the way for the search for new high-performance Heusler compounds.

\section*{Declaration of Competing Interest}

The authors declare that they have no known competing financial interests or personal relationships that could have appeared to influence the work reported in this paper.

\section*{Acknowledgments}

We thank Dr. E. Xiao, Dr. W. Zhou, Dr. Y. Sakuraba and Dr. Y. Iwasaki at the National Institute for Materials Science (NIMS) for fruitful discussion. This work was supported by JST-CREST Grant Number JPMJCR21O1 and 
by MEXT Program: Data Creation and Utilization-Type Material Research and Development Project (Digital Transformation Initiative Center for Magnetic Materials) Grant Number JPMXP1122715503 and by JST-ERATO Grant Number JPMJER2201. The calculations in this study were performed on the Numerical Materials Simulator at NIMS.

\printcredits

\bibliographystyle{elsarticle-num}

\bibliography{Heusler-refs_cured}

\begin{thebibliography}{10}
\bibitem{L21}
The crystal structure of Ni-doped Co$_3$Sn is usually called as DO$_3$, which
  is a sepcial case of $L2_1$.
\end{thebibliography}



\end{document}